# On the $0^{++}$ Glueball Mass

Stuart Samuel

Department of Physics
The City College of New York
New York, New York, 10031 USA

**ABSTRACT**

An approximate vacuum wave functional $\Psi_0$ is proposed for 2+1-dimensional Yang-Mills theories. Using $\Psi_0$, one can compute the $0^{++}$ glueball mass $M_G$ in terms of the string tension. By using the idea of dimensional reduction, a prediction for $M_G$ can be made in 3+1 dimensions. One finds $M_G \approx 1.5$ GeV.

# I. Introduction

Yang-Mills theories without quarks are expected to produce one or more bound states or glueballs. When quarks are present, such pure glue states should mix with $q\bar{q}$ states but some meson might have a dominant gluon content. Possible candidates for such states are $f_0(1300)$, $f_0(1590)$, $f_J(1710)$ and $\eta(1440)$.[1] The $\eta'(958)$ is expected to obtain a large fraction of its mass due to fluctuations in gluonic topological charge.[2-4]

Initial lattice studies suggested that the lightest $0^{++}$ glueball might have a mass $M_G$ of ~1 GeV. Such lattice glueball computations have been among the most numerically demanding due to a poor signal to noise ratio. However, due to advances in algorithms and powerful computers, much progress has been made. The most recent lattice results give values of $M_G$ of $1550 \pm 50$ MeV [5] and $1740 \pm 71$ MeV [6], suggesting that the $f_J(1710)$ or $f_0(1590)$ might be the lightest $0^{++}$ glueball state. The $f_J(1710)$ is favored since it has decay widths consistent with numerical simulations.[7]

The purpose of this letter is to obtain an approximate analytic computation of the $0^{++}$ glueball mass in terms of the string tension $\sigma$. The calculation is carried out in D=2+1 dimensions and extended by dimensional reduction to D=3+1 dimensions. Analytic studies often provide more physical insight than numerical simulations. In addition, the interplay between numerical and analytic approaches can assist either approach in obtaining new methods and results.

Our method consists in postulating the form of the ground state wave functional $\Psi_0$. The proposed $\Psi_0$ is adjusted to agree with weak and strong coupling limits. Support for our $\Psi_0$ comes from the previous work [8,9]. In particular, numerical and analytic studies [10-17] of the lattice Hamiltonian formulation [18] give rise to vacuum functionals similar to our $\Psi_0$.

The Hamiltonian for the D=d+1 dimensional Yang-Mills theory is

$$H = \frac{g^2}{2} \int d^dx \, E_i^a \, E_i^a(x) + \frac{1}{2g^2} \int d^dx \, B_{ij}^a \, B_{ij}^a(x) \quad , \qquad (1.1)$$

where $E_i^a(x) = \frac{1}{i}\frac{\delta}{\delta A_i^a(x)}$, $B_{ij}^a = \partial_i A_j^a - \partial_j A_i^a + f^{abc}A_i^b A_j^c$, g is the gauge coupling constant and $f^{abc}$ are the structure constants: $[\lambda^a, \lambda^b] = if^{abc}\lambda^c$, where the Lie-algebra generators $\lambda^a$ are normalized so that $Tr(\lambda^a \lambda^b) = \delta^{ab}/2$.

In 2+1 dimensions, our key result is

$$M_G \approx \frac{8\sigma}{g^2 C_f} \quad , \tag{1.2}$$

where $C_f$ is the value of casimir operator for the fundamental representation: $\lambda^a \lambda^a = C_f I$. For SU(N), $C_f = (N^2-1)/(2N)$.

## II. The Approximate Vacuum Wave Functional

Express the ground state $\Psi_0$ as

$$\Psi_0 = \exp(-f(B)) \quad , \tag{2.1}$$

where f is a functional of B. As g goes to zero, the perturbative $f_{pt}(B)$ is

$$f_{pt}(B) = \frac{1}{4g^2}\int d^d x B_{ij}^a(x)\left(\frac{1}{\sqrt{-\partial^2}}B_{ij}^a(x)\right)$$

$$= \frac{1}{4g^2}\int d^d x \int d^d y B_{ij}^a(x)\left(\frac{1}{\sqrt{-\partial^2}}\right)(x,y)B_{ij}^a(y) \quad , \tag{2.2}$$

where the kernel $(-\partial^2)^{-1/2}(x,y)$ is

$$\left(\frac{1}{\sqrt{-\partial^2}}\right)(x,y) = \int \frac{d^d p}{(2\pi)^d}\frac{\exp(ip \cdot (x-y))}{\sqrt{p \cdot p}} \quad . \tag{2.3}$$

In the abelian case for which there are no interactions, perturbation theory and Eq.(2.2) are exact. For the non-abelian case, Eq.(2.2) is not gauge-invariant but the violations of gauge invariance are of order $g^2$.

It has been conjectured that, in strong coupling, the ground state wave functional is governed by $f_{sc}(B)$ with [9]

$$f_{sc}(B) = \frac{\mu}{2}\int d^d x\, B_{ij}^a\, B_{ij}^a(x) \quad , \tag{2.4}$$

where μ is a parameter with dimensions of length$^{(4-d)}$. There is much evidence in support of Eq.(2.4).[9-17] In particular, lattice theory gives Eq.(2.4) as the leading strong coupling result.[15]

In a confining theory in space-time dimensions D with $2 < D \leq 4$, Eq.(2.4) can be derived by the following argument. Work in a basis of generalized closed Wilson loops of arbitrary shape and number and in arbitrary representations. This gauge-invariant set forms a complete set of variables. Expand the vacuum functional in such a basis. Using constructive field theory, the vacuum state can be obtained by doing a functional integral weighted by exp(-S) in which one integrates over half of space-time corresponding to t<0 and uses free boundary conditions at t=0. The functional at t=0, obtained in this way, is the exact ground state wave functional. In a confining theory, Wilson loops of large area are suppressed. Due to the kinetic energy term $g^2 \int d^dx\, E_i^a E_i^a(x)/2$ in H, one sees that the there is also a contribution per unit length to the energy. Hence, Wilson loops of large area or large perimeter are suppressed. One concludes that, in a confining theory, f(B) in Eq.(2.1) is a sum over arbitrary numbers of Wilson loops in arbitrary representations for loops which are of small size. As a consequence, f(B) acts like a localized field theory. In computing vacuum expectation values, a confining theory in D dimensions becomes a localized field theory in d=D-1 dimensions governed by an action equal to 2f(B) (the factor of two arises because $|\Psi_o|^2$ enters in computations). Now, use the idea behind the renormalization group. To compute the behavior of a large Wilson loop or long-distance correlation function, one may integrate out short-distance degrees of freedom. Integrate out to a scale slightly larger that the confinement distance. Then, the resulting field theory will be dominated by the local gauge-invariant operator of lowest dimension. This operator is $B_{ij}^a B_{ij}^a$. In space-time dimensions D with $D \leq 4$, strong coupling corresponds to large distances. Hence, the effective vacuum functional in the strong coupling limit is given by Eq.(2.4).

From the above discussion, it is clear that the true vacuum functional is quite complicated. However, a simplified functional, which interpolates between the weak coupling (Eq.(2.2)) and strong coupling (Eq.(2.4)) forms, might produce good results for computations. Consider the approximate vacuum functional governed by

$$f(B) = \frac{1}{2g^2}\int d^d x \mathrm{Tr}\left(B_{ij}(x)\left(-\mathcal{D}^k\mathcal{D}_k + m_0^2\right)^{-1/2}(x,y)B_{ij}(y)\right) , \qquad (2.5)$$

where $B_{ij}(x) = \lambda^a B_{ij}^a(x)$, Tr stands for trace, $\mathcal{D}_k$ is the covariant derivative in the adjoint representation and $m_0$ is a mass parameter.

At short distances and small coupling, the derivative term in f(B) dominates and Eq.(2.5) reduces to Eq.(2.2). At large distances, the mass term dominates and f(B) in Eq.(2.5) reduces to Eq.(2.4) with

$$\mu = \frac{1}{2m\ g^2} . \qquad (2.6)$$

Here, we have replaced the parameter $m_0$ by a renormalized parameter m. In integrating out short-distance degrees of freedom, one expects $m_0$ to be renormalized. In particular, m contains contributions related to the energy per unit length of the Wilson lines which enter Eq.(2.5), as we now explain.

An explicit formula for the kernel in Eq.(2.5) for SU(N) is

$$\left(-\mathcal{D}^k\mathcal{D}_k + m_0^2\right)^{-1/2}_{s_1 t_1; s_2 t_2}(x,y) = \int_0^\infty \frac{d\tau}{\sqrt{2\pi\tau}} \times$$

$$\int_{\substack{X(0)=y\\X(\tau)=x}}\left[\prod_\tau \mathcal{D}X(\tau)\right]\exp\left(-\frac{1}{2}\int_0^\tau d\sigma\left(\dot X^2(\sigma) + m_0^2\right)\right) \times \qquad (2.7)$$

$$\left\{\left[\wp_{x\leftarrow y}\exp\left(i\int_0^\tau d\sigma A_i \dot X^i(\sigma)\right)\right]_{t_1 s_2}\left[\wp_{y\leftarrow x}\exp\left(i\int_\tau^0 d\sigma A_i \dot X^i(\sigma)\right)\right]_{t_2 s_1}\right\},$$

where $\wp$ denotes the path-order product along $X(\sigma)$ and $A_i = \lambda^a A_i^a(X(\sigma))$. The measure $\mathcal{D}X$ in Eq.(2.7) is the Feynman one [19] for a particle of unit mass in d dimensions. The subscripts $s_1$, $t_1$, $s_2$, $t_2$, which run from 1 to N for SU(N), are matrix indices:

$$\text{Tr}\left(B_{ij}(x)\left(-\mathcal{D}^k\mathcal{D}_k + m_0^2\right)^{-1/2}(x,y)\ B_{ij}(y)\right) =$$

$$B_{ij}^{s_1 t_1}(x)\left(-\mathcal{D}^k\mathcal{D}_k + m_0^2\right)^{-1/2}_{s_1 t_1; s_2 t_2}(x,y)\ B_{ij}^{s_2 t_2}(y)\ . \qquad (2.8)$$

Eq.(2.7) shows that $\left(-\mathcal{D}^k\mathcal{D}_k + m_0^2\right)^{-1/2}(x,y)$ involves a sum over paths of a Wilson line in the adjoint representation which goes from the space-point y to the space-point x. Each path contributes with a particular probability. As $m_0$ is increased, paths of smaller size are weighted comparatively more than paths of larger size. The Wilson lines in Eq.(2.7) render Eq.(2.5) gauge invariant. When the kinetic term $g^2 \int d^dx\ E_i^a E_i^a(x)/2$ in H acts on the Wilson line, a contribution to the energy proportional to the length of the non-backtracking part of the path is produced. Hence, it is energetically favorable to have $m_0$ be non-zero. In the case of a U(1) group, the matrix factor in curly brackets in Eq.(2.7) is replaced by 1 and no such Wilson line contribution arises and it is energetically favorable for $m_0$ to be zero.

Besides having the correct strong and weak coupling limits, there is numerical evidence from lattice simulations that $\Psi_0$ in Eqs.(2.1) and (2.5) is a reasonable approximation for the 2+1 dimensional Yang-Mills theory. Consider expanding f(B) in inverse powers of $m_0$. The result is

$$f(B) = \mu_0 \int d^2x \text{Tr}(B(x)B(x)) + \mu_2 \int d^2x \text{Tr}\left(\mathcal{D}^k B(x) \mathcal{D}_k B(x)\right) + \ldots\ , \qquad (2.9)$$

where

$$\mu_0 = \frac{1}{2m_0 g^2}\ , \qquad \mu_2 = -\frac{1}{4m_0^3 g^2}\ . \qquad (2.10)$$

Ref.[13] has performed Monte Carlo simulations of the ground state functional at intermediate couplings for SU(2) and found that the form in Eq.(2.9) fits the data well with $\mu_0 = (0.91 \pm 0.02)/g^4$ and $\mu_2 = -(0.19 \pm 0.05)/g^8$. Analytic strong coupling lattice computations in the Hamilonian formulation lead to similar results.[16] As predicted from Eq.(2.5), $\mu_2$ should be negative and this is borne out in simulations. The Monte Carlo data implies $m_0 \approx 1.55\ g^2$ for SU(2).

## III. Consequences of The Approximate Vacuum Functional

Let us first consider D=2+1 space-time dimensions. In this case, the strong-coupling functional in Eq.(2.5) leads to confinement because the dimensionally reduced effective theory is similar to a localized Yang-Mills theory in two dimensions. The string tension $\sigma$ is obtained from the vacuum expectation value of a spatially oriented Wilson loop. The result is

$$\sigma = \frac{g^2 m C_f}{4} \quad . \tag{3.1}$$

The $0^{++}$ glueball mass $M_G$ is obtained as the coefficient of the exponential falloff of the correlation function $<\frac{1}{2} B^a_{ij} B^a_{ij}(0,0) \; \frac{1}{2} B^a_{kl} B^a_{kl}(0,t)> \sim c(t) \exp(-M_G t)$ for large t. By exploiting Lorentz invariance, $M_G$ can be extracted from the vacuum functional via

$$<\Psi_0 | \tfrac{1}{2} B^a_{ij} B^a_{ij}(x) \; \tfrac{1}{2} B^a_{kl} B^a_{kl}(y) | \Psi_0> \sim c(r) \exp(-M_G r) \quad , \tag{3.2}$$

where $r = |x - y|$. In two space dimensions and at large distances, the propagator for $B^a_{12}$ in the effective field theory governed by $|\Psi_0|^2$ is $(-\partial^2 + m^2)^{1/2}$. One finds, using perturbation theory in this effective theory, that Eq.(3.2) holds with

$$M_G \approx 2m \quad . \tag{3.3}$$

Eq.(3.3) has a physical interpretation. The parameter m can be thought of as a constituent mass for a gluon. More precisely, m is the effective mass of an adjoint-representation configuration of gluons inside a bound state. Since two adjoint representations are needed to make a singlet, the glueball mass is approximately 2m.

Combining (3.1) and (3.3), one arrives at the main result in Eq.(1.2).

Monte Carlo simulations of the 2+1 dimensional SU(2) Yang-Mills theory have accurately [20] determined the string tension to be $\sigma_{MC} = (0.112 \pm 0.002) g^4$. Using this value for $\sigma$ in Eq.(1.2), we obtain $M_G \approx 1.2 \, g^2$. The Monte Carlo calculations of ref.[21] give $M_G = (1.59 \pm 0.01) g^2$. Thus, we find $M_G/\sqrt{\sigma} \approx 3.6$, while numerical simulations [21] give $M_G/\sqrt{\sigma} = 4.77 \pm 0.05$. Using the difference between our results and those of the lattice as a means of estimating systematic uncertainty, our method for computing the $0^{++}$ glueball mass is accurate to about 25%. For a summary of computations of $M_G/\sqrt{\sigma}$, see the references in ref.[22].

For SU(3), Monte Carlo simulations give $\sigma_{MC} = (0.307 \pm 0.004)\, g^4$.[23] This value of $\sigma$ in conjunction with Eq.(1.2) leads to the SU(3) results

$$M_G \approx 1.84\, g^2\, , \qquad M_G/\sqrt{\sigma} \approx 3.3\ . \tag{3.4}$$

The result in Eq.(3.4) is in reasonably agreement with strong coupling Hamiltonian computations [24].

Recently, another analytic approach, based on finding exact eigenstates of the kinetic energy term in H, gives $M_G = N/\pi\, g^2$ for SU(N).[25] For SU(2), one obtains $M_G \approx 0.64\, g^2$, which is considerably smaller than the Monte Carlo result, suggesting that corrections involving the potential energy term $\int d^2x\, B^a_{12}\, B^a_{12}(x)/2g^2$ may be important. For SU(3), the approach gives $M_G \approx 0.95\, g^2$, which is about half the value in Eq.(3.4).

## IV. Extrapolation to D=4

The result in Eq.(1.2) does not directly apply to D=3+1 dimensions. However, one can appeal to the idea of dimensional reduction.[26] If a 3+1 dimensional gauge theory is confining, then the computation of spatial correlation functions and Wilson loops involves only those degrees of freedom within the order of $L_c$ in the time direction, where $L_c$ is the confinement length. Hence, it should be possible to approximate a 3+1 dimensional confining theory by a 2+1 dimensional theory. This idea has been used several times in past work[15,27] and is embodied in the result that $f_{sc}(B)$ can be used for the purposes of computing large spatial Wilson loops. Due to the beta function being negative, $\mu$ is expected to be exponentially large in $1/g^2$. In short, assuming four-dimensional confinement, results for $M_G/\sqrt{\sigma}$ should be approximately the same in 2+1 and 3+1 dimensions. In fact, Monte Carlo simulations of the SU(2) gauge theory indicate that the glueball spectra, in units of $\sqrt{\sigma}$, in 2+1 and 3+1 dimensions agree to about 15%.[21]

Assuming the validity of dimensional reduction, Eq.(3.4) with $\sqrt{\sigma}$ = 440 MeV gives a value of the $0^{++}$ glueball mass $M_G$ in D=3+1 dimensions for the SU(3) gauge theory of

$$M_G \approx 1.5 \text{ GeV} \quad . \tag{4.1}$$

## V. Summary

In summary, by using an approximate vacuum functional which interpolates between strong and weak coupling forms, we have obtained the relation in Eq.(1.2) between the lightest $0^{++}$ glueball mass and the string tension in 2+1 dimensions. Using dimensional reduction, we obtain a value of about 1.5 GeV for $M_G$ in 3+1 dimensions.

Our value of 1.5 GeV for $M_G$ is much larger than many of the natural scales in Yang-Mills theory: $\sqrt{\sigma}$ is 440 MeV, the deconfining phase transition temperature is ~250 MeV, the mass of the color-singlet gluon cloud around a quark is ~300 MeV assuming that the contribution to the constituent mass of a quark in a bound state comes from such a gluon cloud, and the topological susceptibility $\langle \nu^2 \rangle$ which enters the $\eta'$ mass is $\langle \nu^2 \rangle^{1/4} \approx 180$ MeV [28-33]. On this basis one might expect approximate analytic computations of $M_G$ to yield values less than 1 GeV. For example, if the continuum strong coupling 2+1 dimensional Hamiltonian result [25] is assumed to extrapolate to D=3+1 by dimensional reduction, one would obtain $M_G \approx 760$ MeV.

Since, as mentioned above, the error in our results in 2+1 dimensions is estimated to be about 25% and that the error in extrapolating to 3+1 dimensions is about 15%, the total error in $M_G$ is around 30%. Our value of 1.5 GeV for $M_G$ is about 15% less than the most accurate current lattice value of 1740 MeV [6].

**Acknowledgments**

We thank V. P. Nair for discussions and Columbia University for hospitality. This work is supported in part by the United States Department of Energy under the grant number DE-FG02-92ER40698.

# References

[1] Particle Data Group, Phys. Rev. **D50** (1994) 1173.

[2] H. Goldberg, Phys. Rev. Lett. **44** (1980) 363.

[3] E. Witten, Nucl. Phys. **B156** (1979) 269.

[4] G. Veneziano, Nucl. Phys. **B159** (1979) 213.

[5] G. Bali, K. Schilling, A. Hulsebos, A. Irving, C. Michael and P. Stephenson, Phys. Lett. **B309** (1993) 378.

[6] J. Sexton, A. Vaccarino and D. Weingarten, Nucl. Phys. **B** (Proc. Suppl.) **34** (1994) 357.

[7] J. Sexton, A. Vaccarino and D. Weingarten, *Numerical Evidence for the Observation of a Scalar Glueball*, preprint IBM-HET-95-2.

[8] R. P. Feynman, Nucl. Phys. **B188** (1981) 479.

[9] J. Greensite, Nucl. Phys. **B158** (1979) 469.

[10] J. Greensite, Phys. Lett. **B191** (1987) 431.

[11] J. Greensite and J. Iwasaki, Phys. Lett. **B223** (1989) 207.

[12] H. Arisue, M. Kato and T. Fujiwara, Prog. Theor. Phys. **70** (1983) 229.

[13] H. Arisue, Phys. Lett. **B280** (1992) 85.

[14] H. Arisue, Prog. Theor. Phys. **84** (1990) 951.

[15] J. Greensite, Nucl. Phys. **B166** (1980) 113.

[16] S.-H. Guo, Q.-Z. Chen and L. Li, Phys. Rev. **D49** (1994) 507.

[17] Q.-Z. Chen, X.-Q. Luo and S.-H. Guo, Phys. Lett. **B341** (1995) 349.

[18] J. Kogut and L. Susskind, Phys. Rev. **D11** (1975) 395.

[19] R. P. Feynman and A. R. Hibbs, *Quantum Mechanics and Path Integrals*, (McGraw-Hill, 1965).

[20]  M. Teper, Phys. Lett. **B311** (1993) 223.

[21]  M. Teper, Phys. Lett. **B289** (1992) 115;
Nucl. Phys. **B** (Proc. Suppl.) **30** (1993) 529.

[22]  C. J. Hamer, M. Sheppeard, Z. Weihong, and D. Schütte, *Universality for SU(2) Yang-Mills Theory in (2+1)D*, preprint hep-th/9511179.

[23]  M. Lütgemeier, Nucl. Phys. **B** (Proc. Suppl.) **42** (1995) 523.

[24]  Q. Z. Chen, X. Q. Luo, S. H. Guo and X. Y. Fang, Phys. Lett. **B348** (1995) 560.

[25]  D. Karabali and V. P. Nair, *A Gauge-Invariant Hamiltonian Analysis for Non-Abelian Gauge Theories in (2+1) Dimensions*, preprint CCNY-HEP 95/6 (October 1995).

[26]  S. Samuel, Nucl. Phys. **B154** (1979) 62.

[27]  J. Ambjørn, P. Olesen and C. Peterson, Nucl. Phys. **B240** (1994) 189.

[28]  P. Woit, Phys. Rev. Lett. **51** (1983) 638.

[29]  J. Hoek, M. Teper and J. Waterhouse, Phys. Lett. **B180** (1986) 112.

[30]  M. Göckeler, A. S. Kronfeld, M. L. Laursen, G. Schierholz and U.-J. Wiese, Nucl. Phys. **B292** (1987) 349.

[31]  A. S. Kronfeld, M. L. Laursen, G. Schierholz, S. Schleiermacher and U.-J. Wiese, Nucl. Phys. **B305** (1988) 661.

[32]  M. Teper, Phys. Lett. **B202** (1988) 553.

[33]  M. Campostrini, A. Di Giacomo, H. Panagopoulos and E. Vicari, Nucl. Phys. **B329** (1990) 683.